\documentclass[prl,
 reprint,
superscriptaddress,
preprintnumbers,
nofootinbib,
 amsmath,amssymb,
 aps,
]{revtex4-1}
\usepackage[ocgcolorlinks]{hyperref}

\usepackage[utf8]{inputenc}

\newcommand{\GeV}{\text{\,GeV}}

\newcommand{\be}{\begin{equation}}
\newcommand{\ee}{\end{equation}}
\newcommand{\ba}{\begin{eqnarray}}
\newcommand{\ea}{\end{eqnarray}}

\begin{document}

\preprint{INR-TH-2020-033}

\title{No Miracle in Gravity Portals} 

\author{Fedor Bezrukov}
\email{Fedor.Bezrukov@manchester.ac.uk}
\affiliation{The University of Manchester, Department of Physics and Astronomy, Oxford Road, Manchester M13 9PL, United Kingdom}
\author{Sergey Demidov}
\email{demidov@ms2.inr.ac.ru}
\affiliation{Institute for Nuclear Research of the Russian Academy of Sciences, 117312 Moscow, Russia}
\affiliation{Moscow Institute of Physics and Technology, 141700 Dolgoprudny, Russia}
\author{Dmitry Gorbunov}
\email{gorby@ms2.inr.ac.ru}
\affiliation{Institute for Nuclear Research of the Russian Academy of Sciences, 117312 Moscow, Russia}
\affiliation{Moscow Institute of Physics and Technology, 141700 Dolgoprudny, Russia}

\date{\today}

\begin{abstract}
  The idea of dark matter particles coupled only gravitationally is minimalist yet viable. Assuming an additional $Z_2$-breaking linear coupling of scalar curvature to the dark matter scalar (gravity portal) Refs.~\cite{Cata:2016dsg,Cata:2016epa} claimed a strong parametric growth of the dark matter particle decay rate with its mass, which implies pronounced phenomenological signatures for the model. This peculiarity was attributed by the authors to the enhancement due to the presence of longitudinal gauge bosons in the final state. Quite unfortunately there were overlooked cancellations in the tree-level amplitudes. There is no miracle: all perturbative decay rates are suppressed by the strong coupling scale.   
\end{abstract}

\maketitle

For a free scalar particle $\phi$ of mass $m_\phi$, playing the role of dark matter, a $Z_2$-violating minimal coupling to gravity via Ricci scalar $R$ is written in Refs.~\cite{Cata:2016dsg,Cata:2016epa} as
\begin{equation}
  \label{GP}
  \mathcal{L}_\xi = -\xi M R \phi,    
\end{equation}
involving dimensionless parameter $\xi$ and mass scale $M$. Naturally this term induces decay of the otherwise stable scalar $\phi$ to gravitons with decay rate suppressed by the perturbative strong coupling scale $\Lambda\equiv M_P^2/\xi M$
\begin{equation}
  \label{decay-GG}
  \Gamma_{\phi\to GG} \sim \frac{m_\phi^3}{\Lambda^2} = \frac{\xi^2 M^2 m_\phi^3}{M^4_{P}},   
\end{equation}
where $M_P=2.4\times 10^{18}\GeV$ is the reduced Planck mass. Likewise, this coupling induces the decay of $\phi$ to all other particles in the entire model with gravity playing the role of a messenger, since all particles participate in gravitational interactions. The resulting effective interactions become explicit upon the Weyl transformation of the metric $g_{\mu\nu}$ to that with the Einstein--Hilbert gravity
\[
  \tilde g_{\mu\nu} = \Omega^2 g_{\mu\nu}, \quad \Omega^2 = 1+2\xi M \phi/M_P^2, 
\]
and conformal rescaling of the fermionic fields  $\tilde{\psi}_f=\Omega^{-3/2}\psi_f$, leaving us with the interaction Lagrangian
\begin{equation}
  \label{int}
  \mathcal{L}_{\text{int}} = \Omega^{-2}\tilde T_H + \Omega^{-1}\tilde L_Y - \Omega^{-4}V_H.
\end{equation}
Here $\tilde T_H$ stands for the kinetic term for the Higgs field, and $\tilde L_Y$ and $V_H$ for the Yukawa terms and the Higgs potential, respectively.  As compared to \cite{Cata:2016dsg,Cata:2016epa} the kinetic term for fermions is absent, which corresponds to the fact that it is conformally invariant \cite{Birrell:1982ix}, and thus cancellations in the fermionic sector written in terms of not rescaled field $\psi_f$ are automatic.

The Yukawa and potential terms in \eqref{int} lead to the decays of the scalar to Standard Model (SM) particles additionally suppressed by SM couplings, while the interaction with the Higgs kinetic term gives rise to $\phi\to hh$ decay with behaviour similar to \eqref{decay-GG}. However, Refs.~\cite{Cata:2016dsg,Cata:2016epa} claimed that while this is typically the case, for decays with SM massive gauge vector bosons, $W^\pm$, $Z$ in the final state the rates get amplified by the powers of $m_\phi^2/v^2$ where $v=246$\,GeV is the SM Higgs field  vacuum expectation value.  This finding changes dramatically all the phenomenology of the heavy scalar $\phi$ and has attracted some interest in literature to the gravity portal \eqref{GP}. This result is extremely surprising, given that the Goldstone boson equivalence theorem \cite{Cornwall:1974km,Vayonakis:1976vz} predicts that these widths should behave as the widths into Goldstone bosons from the first term of \eqref{int}, which have the form \eqref{decay-GG}.  Moreover, such amplification would mean that the theory looses unitarity in the symmetry restoring limit $v\to0$.

Inspired by these ideas we decided to verify the results presented in Refs.~\cite{Cata:2016dsg,Cata:2016epa}, and found that \emph{none} of the decay rates of $\phi$ is amplified by $m_\phi^2/v^2$ factor.

For conformally rescaled fermions the $\phi\bar f f V$-vertex in Appendix A1 of Ref.~\cite{Cata:2016epa} is absent. Thus the decay into vector boson and fermion pair $\phi\to f\bar f V$ can proceed only via intermediate vector boson $\phi\to VV\to V f\bar{f}$ or fermion $\phi\to f\bar f\to V f \bar f$ and width \eqref{decay-GG} is additionally suppressed by the SM couplings constants.

For the interaction in the inflaton-Higgs-gauge boson sector required for $\phi\to HVV$ and $\phi\to HHVV$ decays all the relevant Feynman rules in Appendix A1 of Ref.~\cite{Cata:2016epa} are correct, but the leading order contributions of longitudinal vector modes enhanced by factors of momenta to the vector boson mass ratio cancel exactly in the corresponding squared matrix elements. Independently, we checked this with both analytical and numerical calculations provided within the {LanHEP}~\cite{Semenov:1996es,Semenov:2014rea} and {CompHEP} \cite{Pukhov:1999gg,Boos:2004kh}/\allowbreak{CalcHEP}~\cite{Belyaev:2012qa} packages. Therefore, we conclude that the longitudinal boson equivalence theorem \cite{Cornwall:1974km,Vayonakis:1976vz} holds, and there are no amplifications of the scalar decay rates by $m_\phi^2/v^2$ with respect to \eqref{decay-GG}.   

\begin{acknowledgments}
The analytic calculations are supported by the Russian Science Foundation grant 19-12-00393. The 
simulations with {LanHEP}/\allowbreak{CalcHEP} packages are supported by the STFC research grant ST/P000800/1.
\end{acknowledgments}

\bibliography{GP}

\end{document}